\documentclass[10pt,conference]{IEEEtran}\IEEEoverridecommandlockouts
\usepackage{epsfig,graphicx,subfigure,psfrag,amsmath,cases}
\usepackage{latexsym,amssymb,epsfig,subfigure,algorithm}
\usepackage{algorithmic}
\usepackage{color}
\usepackage{url}
\usepackage{scrtime}
\usepackage{bm}
\usepackage{framed}
\author{\authorblockN{Elena Boshkovska, Derrick Wing Kwan Ng, Nikola Zlatanov, and Robert Schober}\vspace*{-1.0cm}}

\title{\vspace*{-0.5cm}Practical Non-linear Energy Harvesting Model and Resource Allocation for SWIPT Systems\vspace*{-0.3cm}\vspace*{-0.0cm}}

\date{\thistime,\,\today}

\newtheorem{Thm}{Theorem}

\newtheorem{T-Prob}{Transformed Problem}

\DeclareMathOperator{\Tr}{\mathrm{Tr}}
\DeclareMathOperator{\zero}{\mathbf{0}}
\DeclareMathOperator{\Rank}{\mathrm{Rank}}

\DeclareMathOperator{\maxo}{\mathrm{maximize}}
\DeclareMathOperator{\mino}{\mathrm{minimize}}

 \newcommand{\qed}{\hfill \ensuremath{\blacksquare}}
\newtheorem{Remark}{Remark}

\newcommand{\abs}[1]{\lvert#1\rvert}

\newcommand{\norm}[1]{\lVert#1\rVert}

\begin{document}

\maketitle

\begin{abstract}
In this letter, we propose a practical non-linear energy harvesting model and design a resource allocation algorithm for
 simultaneous wireless information and power transfer (SWIPT) systems. The algorithm design is formulated as a
non-convex optimization problem for the maximization of the total  harvested power at energy harvesting receivers
 subject to  minimum required signal-to-interference-plus-noise ratios (SINRs) at multiple information receivers. We transform  the considered non-convex objective function from sum-of-ratios form into an equivalent objective function in subtractive
form, which enables the derivation of an efficient iterative
resource allocation algorithm.  In each iteration, a rank-constrained semidefinite program (SDP) is solved optimally by SDP relaxation.   Numerical results unveil a substantial performance gain that can be achieved if the resource allocation design is based on the proposed non-linear energy harvesting model instead of the traditional linear model.
\end{abstract}
\renewcommand{\baselinestretch}{0.90}
\normalsize
\section{Introduction}
Energy harvesting (EH) is a promising solution for prolonging the lifetime of communication networks by introducing self-sustainability to energy-limited devices. Among different EH technologies, wireless power transfer (WPT) via electromagnetic waves in radio frequency (RF) enables comparatively controllable EH at the receivers compared to conventional natural energy sources, such as wind and solar. Recent progress in the development of RF-EH circuitries has made RF-EH practical for  low-power consumption devices, e.g. wireless sensors. In particular, RF-EH provides the possibility of simultaneous wireless information and power transfer (SWIPT) \cite{Krikidis2014}\nocite{JR:Xiaoming_magazine,JR:MIMO_WIPT,JR:WIPT_fullpaper_OFDMA}--\cite{JR:EE_SWIPT_Massive_MIMO}. Yet, this new technology introduces a paradigm shift in system and resource allocation algorithm design. In \cite{JR:MIMO_WIPT}, the authors studied rate-energy trade-off regions by designing an optimal beamformer.
In \cite{JR:WIPT_fullpaper_OFDMA}, energy-efficient SWIPT was investigated in multicarrier systems, where power allocation, user scheduling, and subcarrier allocation were considered.  In \cite{JR:EE_SWIPT_Massive_MIMO}, the authors solved the energy efficiency maximization problem for large-scale multiple-antenna SWIPT systems. However, existing literature studies \cite{Krikidis2014} and resource allocation algorithm designs  for SWIPT networks \cite{JR:MIMO_WIPT}\nocite{JR:WIPT_fullpaper_OFDMA}--\cite{JR:EE_SWIPT_Massive_MIMO}  are based on a linear EH model where the  RF-to-direct current (DC) power conversion efficiency is independent of the input power level of the EH circuit. { In practice,  EH circuits \cite{JR:Energy_harvesting_circuit}\nocite{JR:EH_measurement_1}--\cite{CN:EH_measurement_2} usually result in a non-linear end-to-end wireless power transfer.} Hence,
the conventional linear EH model cannot properly model the power dependent EH efficiency which leads to a mismatch for resource allocation. To the best of the authors' knowledge, a practical non-linear EH model and a corresponding resource allocation algorithm design for SWIPT networks has not been reported in the literature, yet.

In this letter, we address the above issues. To this end, we first propose a practical parametric non-linear EH harvesting model and verify its accuracy with measurement data. Then, we formulate the resource allocation algorithm design as a non-convex optimization problem for maximization of the total harvested energy. The considered non-convex optimization problem is solved optimally by an iterative algorithm. Simulation results illustrate the total harvested energy loss when a conventional linear EH model is adopted for resource allocation algorithm design.

\textbf{Notation:}
$\mathbf{A}^H$, $\Tr(\mathbf{A})$, $\mathbf{A}^{-1}$, and $\Rank(\mathbf{A})$ represent the  Hermitian transpose, trace, inverse, and rank of  matrix $\mathbf{A}$, respectively; $\mathbf{A}\succeq \mathbf{0}$ indicates that $\mathbf{A}$ is a  positive semidefinite matrix; matrix $\mathbf{I}_{N}$
denotes the $N\times N$ identity matrix.  $\mathbb{C}^{N\times M}$ denotes the space of all $N\times M$ matrices with complex entries.
$\mathbb{H}^N$ represents the set of all $N$-by-$N$ complex Hermitian matrices.
The distribution of a circularly symmetric complex Gaussian (CSCG)
vector with mean vector $\mathbf{x}$ and covariance matrix
$\mathbf{\Sigma}$  is denoted by ${\cal
CN}(\mathbf{x},\mathbf{\Sigma})$, and $\sim$ means ``distributed
as".  $\cal E\{\cdot\}$ denotes statistical expectation.
\section{System Model}
\begin{figure}[t]
\centering
\includegraphics[width=2.2 in]{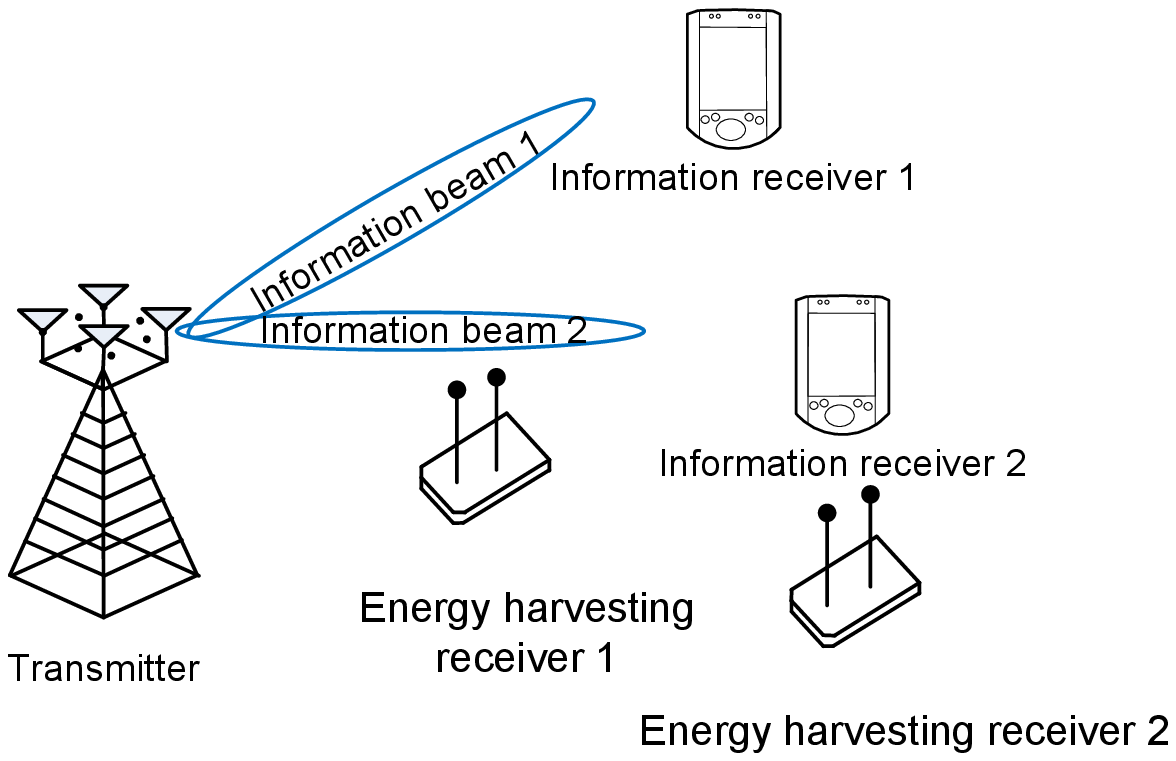}
\caption{A downlink SWIPT system with $K=2$ information receivers (IRs) and $J=2$ energy harvesting receivers (ERs).}
\label{fig:system_model}\vspace*{-4mm}
\end{figure}
\subsection{Channel Model}
We focus on a frequency flat slow fading channel for downlink  multiuser SWIPT systems, cf. Figure \ref{fig:system_model}.   In particular, a transmitter equipped with   $N_\mathrm{T}>1$ antennas serving  $K$  information receivers (IRs) and $J$ energy harvesting receivers (ERs) is considered. The $K$  IRs are low  complexity single-antenna devices and each  ER is equipped with $N_{\mathrm{R}}$ receive antennas to facilitate EH. In each time slot, the transmitter sends a vector of data symbols to the $K$ IRs.  The received signals at IR $k$ and  ER $j$ are given by
\begin{eqnarray}
y_{k}&=&\mathbf{h}_k^H\sum_{k=1}^K\mathbf{w}_k s_k+n_k,\,\,  \forall k\in\{1,\dots,K\},\,\mbox{and}\\
\mathbf{y}_{\mathrm{ER}_j}&=&\mathbf{G}_j^H\sum_{k=1}^K\mathbf{w}_k s_k+\mathbf{n}_{\mathrm{ER}_j},\,\,  \forall j\in\{1,\dots,J\},
\end{eqnarray}respectively, where $s_k\in\mathbb{C}$ and $\mathbf{w}_k\in\mathbb{C}^{N_{\mathrm{T}}\times1}$  are the data symbol  and  the beamforming vector intended for IR $k$, respectively. Without loss of generality, we assume that ${\cal E}\{\abs{s_k}^2\}=1,\forall k\in\{1,\ldots,K\}$. The channel vector between the transmitter and  IR $k$ is denoted by $\mathbf{h}_k\in\mathbb{C}^{N_{\mathrm{T}}\times1}$, and the channel matrix between the transmitter and  ER $j$ is denoted by $\mathbf{G}_j\in\mathbb{C}^{N_{\mathrm{T}}\times N_{\mathrm{R}}}$. $n_k\sim{\cal CN}(0,\sigma_{\mathrm{s}}^2)$ and $\mathbf{n}_{\mathrm{ER}_j}\sim{\cal CN}(\zero,\sigma_{\mathrm{s}}^2\mathbf{I}_{N_{\mathrm{R}}})$ are the additive white Gaussian noises (AWGN) at  the  IRs and the ERs, respectively. $\sigma_{\mathrm{s}}^2$ denotes the noise power at the receiver.

\subsection{Energy Harvesting Model}
In the literature,  the total harvested energy at ER $j$, $\Phi_{\mathrm{ER}_j}^{\mathrm{Linear}}$,  is typically modelled by the following linear model \cite{Krikidis2014}--\cite{JR:EE_SWIPT_Massive_MIMO}:
\begin{eqnarray}\label{eqn:linear_model}
\Phi_{\mathrm{ER}_j}^{\mathrm{Linear}}=\eta_j P_{\mathrm{ER}_j},\quad
P_{\mathrm{ER}_j}=\sum_{k=1}^K\Tr\Big(\mathbf{w}_k\mathbf{w}^H_k\mathbf{G}_j\mathbf{G}_j^H\Big),
\end{eqnarray}
where $P_{\mathrm{ER}_j}$ is the received RF power at ER $j$ and $0\leq\eta_j\leq1$ is the fixed energy conversion efficiency of ER $j$. We note that in this linear EH model, the energy conversion efficiency is independent of the input power level at the ER. In other words, the total harvested energy at the ER is linearly and directly proportional to the received RF power.
{  However, in
practice, EH circuits \cite{JR:Energy_harvesting_circuit}\nocite{JR:EH_measurement_1}--\cite{CN:EH_measurement_2} result in a non-linear end-to-end wireless power transfer.}

In general, for low power, the RF energy conversion efficiency improves as the input power rises, but there are diminishing returns and limitations on
the maximum possible harvested energy, as was evidently proved by field measurements \cite{JR:Energy_harvesting_circuit}\nocite{JR:EH_measurement_1}--\cite{CN:EH_measurement_2}. Thus, it is expected that the conventional linear EH model is only accurate for the specific scenario when the received powers at all ERs are constant. In this letter, we propose a practical  parametric non-linear EH model which captures the dynamics of the RF energy conversion efficiency for different input power levels.  In order to isolate the system model from the specific implementation details of the EH circuit,  we propose a non-linear EH model based on the logistic (sigmoidal) function. Hence, the total harvested energy at ER $j$, $\Phi_{\mathrm{ER}_j}^{\mathrm{Practical}}$, is modelled as:
 \begin{eqnarray}\label{eqn:EH_non_linear}
 \hspace*{-5mm}\Phi_{\mathrm{ER}_j}^{\mathrm{Practical}}\hspace*{-2mm}&=&\hspace*{-2mm}
 \frac{[\Psi_{\mathrm{ER}_j}^{\mathrm{Practical}}
 \hspace*{-0.5mm}- \hspace*{-0.5mm}M_j\Omega_j]}{1-\Omega_j},\, \Omega_j=\frac{1}{1+\exp(a_jb_j)},\\
 \hspace*{-5mm}\Psi_{\mathrm{ER}_j}^{\mathrm{Practical}}\hspace*{-2mm}&=&\hspace*{-2mm} \frac{M_j}{1+\exp\Big(\hspace*{-0.5mm}-a_j(P_{\mathrm{ER}_j}-\hspace*{-0.5mm}b_j)\Big)}.
  \end{eqnarray}
Here, $\Psi_{\mathrm{ER}_j}^{\mathrm{Practical}}$ is the traditional logistic function with respect to the received RF power $P_{\mathrm{ER}_j}$. We introduce a constant $\Omega_j$ in \eqref{eqn:EH_non_linear} to ensure a zero-input/zero-output response for EH.
$M_j$ is a constant denoting the maximum harvested power at ER $j$ when the EH circuit is saturated. Parameters $a_j$ and $b_j$ are constants related to the detailed circuit specifications such as the resistance, capacitance, and diode turn-on voltage. In practice, the EH hardware circuit of each ER is fixed and the parameters $a_j$, $b_j$, and $M_j$ of the proposed model in \eqref{eqn:EH_non_linear}
can be easily found by a standard curve fitting tool. {  We note that the proposed non-linear EH model is able to capture the joint effect of the non-linear phenomena caused by hardware constraints including circuit sensitivity limitations and current leakage \cite{JR:EH_measurement_1,CN:EH_measurement_2}.}

\begin{figure}[t]
\centering\vspace*{-4mm}
\includegraphics[width=3.0 in]{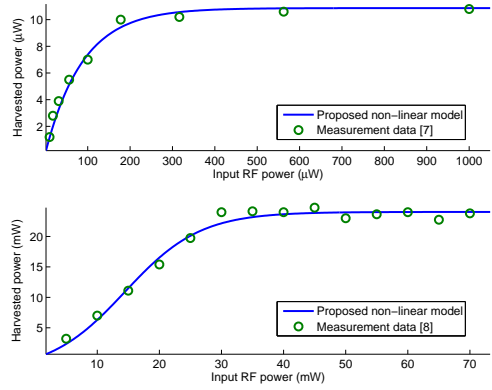}\vspace*{-2mm}
\caption{A comparison between the harvested power for the proposed  model in \eqref{eqn:EH_non_linear} and the measurement data from two different practical EH circuits with different dynamic ranges from \cite{JR:EH_measurement_1} and \cite{CN:EH_measurement_2}.  The parameters $a_j$, $b_j$, and $M_j$ in \eqref{eqn:EH_non_linear} are calculated by a standard curve fitting tool.}
\label{fig:comparsion_measurment}
\end{figure}

{Figure \ref{fig:comparsion_measurment} illustrates that the proposed non-linear EH model closely matches experimental results \cite{JR:EH_measurement_1,CN:EH_measurement_2} for the wireless power harvested by practical EH circuits. Besides, Figure \ref{fig:comparsion_measurment} also shows that the linear model in \eqref{eqn:linear_model} is not accurate in modelling non-linear EH circuits.}

In the sequel, we adopt the proposed non-linear EH model for resource allocation algorithm design. We assume that perfect channel state information is available for resource allocation\footnote{{ In practice, the channel state information of the IRs/ERs can be obtained during the handshaking  between the
transmitter and the receivers before power and data transfer start.}}. Furthermore, since $\Omega_j$ does not affect the design of beamforming vector $\mathbf{w}_k$, cf. \eqref{eqn:EH_non_linear},  for simplicity, we will directly use $\Psi_{\mathrm{ER}_j}^{\mathrm{Practical}}$ to represent the harvested power at ER $j$ in the following study.

\section{Problem Formulation and Solution}
The system design objective is the maximization of the total harvested power and can be mathematically formulated as:
\begin{eqnarray}\label{eqn:TP_maximization}
\underset{\mathbf{w}_{k}}{\maxo}\,\, \hspace*{-2mm}&& \sum_{j=1}^J\Psi_{\mathrm{ER}_j}^{\mathrm{Practical}}\\
\hspace*{-2mm}\mathrm{subject\,\,to}\,\, &&\mathrm{C1}:\,\,\sum_{k=1}^K\norm{\mathbf{w}_k}^2\leq P_{\mathrm{max}},\notag\\
\hspace*{-2mm}&&\mathrm{C2}:\,\, \frac{\mathbf{w}^H_k\mathbf{H}_k\mathbf{w}_k}
{\sum_{j\ne k} \mathbf{w}_j^H\mathbf{H}_k\mathbf{w}_j+\sigma_{\mathrm{s}}^2} \geq \Gamma^{\mathrm{req}}_k,\forall k,\notag
\end{eqnarray}
where $\mathbf{H}_k=\mathbf{h}_k\mathbf{h}_k^H$. Constants $P_{\max}$ and $\Gamma^{\mathrm{req}}_k$  in constraints C1 and C2  are the maximum transmit power for the transmitter and the minimum required signal-to-interference-plus-noise ratio (SINR) at IR $k$, respectively. It can be observed that the objective function in \eqref{eqn:TP_maximization} is in the form of sum-of-ratios which is a non-convex function.
 { In the following, we assume that the considered optimization problem is feasible for the study of resource allocation algorithm design.} Although the Dinkelbach method \cite{JR:Qingqing} or the Charnes--Cooper transformation  can be exploited to handle a single-ratio objective function, they cannot be applied for a sum-of-ratios objective function. In order to obtain a tractable solution, we first transform the non-convex objective function into an equivalent objective function\footnote{Here, ``equivalent" means that the optimization problem with the transformed objective function leads to the same
resource allocation policy as the original problem.} in subtractive form via the following theorem.


\begin{Thm}\label{Thm:1}
Suppose $\mathbf{w}^*_k$ is the optimal solution to   \eqref{eqn:TP_maximization}, then there exist two vectors ${\bm \mu}^*=[\mu_1^*,\ldots,\mu_J^*]$ and ${\bm \beta}^*=[\beta_1^*,\ldots,\beta_J^*]$ such that $\mathbf{w}^*_k$ is an optimal solution to the following optimization problem
\begin{eqnarray} \label{eqn:transformed}
\underset{\mathbf{w}^*_k\in {\cal F}}\maxo\,\sum_{j=1}^J \mu_j^*\Big[\hspace*{-0.5mm}M_j\hspace*{-0.5mm}- \hspace*{-0.5mm} \beta_j^*\Big(1+\exp\big(\hspace*{-0.5mm}-\hspace*{-0.5mm}a_j(P_{\mathrm{ER}_j}\hspace*{-0.5mm}-\hspace*{-0.5mm}b_j)\big)\Big)\hspace*{-0.5mm}\Big],
\end{eqnarray}
where $\cal F$ is the feasible solution set of \eqref{eqn:TP_maximization}. Besides, $\mathbf{w}_k^*$  also satisfies the following system of equations:
\begin{eqnarray} \label{eqn:conditions1}
\beta_j^*\Big(1+\exp\big(\hspace*{-0.5mm}-\hspace*{-0.5mm}a_j(P_{\mathrm{ER}_j}^*\hspace*{-0.5mm}-\hspace*{-0.5mm}b_j)\big)\Big)-M_j&=&0,\\
\mu_j^*\Big(1+\exp\big(\hspace*{-0.5mm}-\hspace*{-0.5mm}a_j(P_{\mathrm{ER}_j}^*\hspace*{-0.5mm}-\hspace*{-0.5mm}b_j)\big)\Big)-1&=&0, \label{eqn:conditions2}
\end{eqnarray}
and $P_{\mathrm{ER}_j}^*=\sum_{k=1}^K\Tr\Big(\mathbf{w}_k^*(\mathbf{w}^*_k)^H\mathbf{G}_j\mathbf{G}_j^H\Big)$.
\end{Thm}

\,\,\emph{Proof:} Please refer to \cite{JR:sum_of_ratios,JR:sum_of_ratios1} for a proof of  Theorem 1.

{
Theorem \ref{Thm:1} suggests that for the maximization problem with sum-of-ratios objective function in \eqref{eqn:TP_maximization}, there exists an equivalent parametric optimization problem with an objective function in subtractive form, such that both problems have the same optimal solution $\mathbf{w}_k^*$. As a result, the optimization problem can be solved by an iterative algorithm consisting of two nested loops. In the inner loop, we solve the optimization in \eqref{eqn:transformed} for given $(\bm{\mu},\bm{\beta})$. Then, in the outer loop,  we find the optimal $(\bm{\mu}^*,\bm{\beta}^*)$ satisfying the system of equations in \eqref{eqn:conditions1} and \eqref{eqn:conditions2}, cf. algorithm in Table \ref{table:algorithm}.}

\begin{table}[t]\caption{Iterative Resource Allocation Algorithm.}\label{table:algorithm}
\vspace*{-5mm}
\begin{algorithm} [H]                    
\renewcommand\thealgorithm{}
\caption{Iterative Resource Allocation Algorithm }          

\label{alg1}                           
\begin{algorithmic} [1]
\STATE Initialize the maximum number of iterations $L_{\max}$, iteration index $n=0$, $\bm\mu$, and $\bm \beta$

\REPEAT [Outer Loop]
\STATE Solve the inner loop problem in \eqref{eqn:rank_constrained} via SDP relaxation for
 given $(\bm\mu^n,\bm\beta^n)$ and obtain the intermediate beamformer $\mathbf{w}_k'$
\IF { \eqref{eqn:convergence_condition} is satisfied} \RETURN
Optimal beamformer $\mathbf{w}_k^*=\mathbf{w}_k'$
 \ELSE \STATE
Update $\bm \mu$ and $\bm\beta$ according to \eqref{eqn:update_beta} and $n=n+1$
 \ENDIF
 \UNTIL{\eqref{eqn:convergence_condition} is satisfied $\,$or $n=L_{\max}$}

\end{algorithmic}
\end{algorithm}\vspace*{-10mm}
\end{table}

\subsection{Solution of the Inner Loop Problem}
 As shown in Table \ref{table:algorithm}, in each iteration in the
inner loop, i.e., in line 3,  we solve the following optimization problem for given
parameters $(\bm{\mu},\bm{\beta})$:
\begin{eqnarray}\label{eqn:rank_constrained}
\underset{\mathbf{W}_k \in\mathbb{H}^{N_{\mathrm{T}}},\tau_j}{\maxo}&&\hspace*{-5mm}\sum_{j=1}^J \mu_j\Big[M_j\hspace*{-0.5mm}- \hspace*{-0.5mm} \beta_j\Big(1+\exp\big(\hspace*{-0.5mm}-\hspace*{-0.5mm}a_j(\tau_j\hspace*{-0.5mm}-\hspace*{-0.5mm}b_j)\big)\Big)\Big] \\
\hspace*{-2mm}\mathrm{subject\,\,to}\,\, &&\hspace*{-5mm}\mathrm{C1}:\,\,\sum_{k=1}^K\Tr(\mathbf{W}_k)\hspace*{-0.5mm} \leq \hspace*{-0.5mm} P_{\mathrm{max}},\notag\\
\hspace*{-2mm}&&\hspace*{-5mm}\mathrm{C2}:\,\, \frac{\Tr(\mathbf{H}_k\mathbf{W}_k)}
{\Gamma^{\mathrm{req}}_k}\hspace*{-0.5mm}  \geq\hspace*{-0.5mm} \sum_{j\ne k} \Tr(\mathbf{H}_k\mathbf{W}_j)+\sigma_{\mathrm{s}}^2,\forall k.\notag\\
\hspace*{-2mm}&&\hspace*{-5mm}\mathrm{C3}:\,\, \Rank(\mathbf{W}_k) \leq\hspace*{-0.5mm} 1,\forall k,\notag\\
\hspace*{-2mm}&&\hspace*{-25mm}\mathrm{C4}:\,\, \tau_j \leq\sum_{k=1}^K\Tr\Big(\mathbf{W}_k\mathbf{G}_j\mathbf{G}_j^H\Big),\forall j,\quad\mathrm{C5}:\,\, \mathbf{W}_k\succeq\zero,\forall k,\notag
\end{eqnarray}
where   $\mathbf{W}_k=\mathbf{w}_k\mathbf{w}_k^H$ and $\tau_j$ are the new  and auxiliary optimization variables, respectively. Although the transformed objective function is in subtractive form,  the transformed optimization problem in \eqref{eqn:rank_constrained}  is still non-convex due to the rank-one matrix constraint C3. To obtain a tractable problem formulation, we apply SDP relaxation. Specifically, we relax constraint C3 in (\ref{eqn:rank_constrained}) by removing constraint $\mathrm{Rank}(\mathbf{W}_k)\leq1$ from the problem. Then, the considered problem becomes a convex SDP problem and can be solved by standard numerical algorithms for convex programs such as the interior point method. Now, we study the tightness of the SDP relaxation in the following theorem.
\begin{Thm}\label{thm:rankone}
Assuming that the channels, i.e., $\mathbf{h}_k$ and $\mathbf{G}_j$,  are statistically independent and \eqref{eqn:rank_constrained} is feasible,
the optimal beamforming matrix of the SDP relaxed problem of (\ref{eqn:rank_constrained}) is a rank-one matrix with probability one, i.e.,  $\Rank(\mathbf{W}_k^*)=1,\forall k,$ for $\Gamma^{\mathrm{req}}_k>0$.
\end{Thm}

\,\,\emph{Proof:} Please refer to the Appendix.

{
Therefore, the adopted SDP relaxation is tight whenever the general channel conditions stated in Theorem $2$ are satisfied. Hence, beamforming is optimal for the maximization of total harvested power for the proposed non-linear EH model. }

\subsection{Solution of the Outer Loop Problem}
In this section, we present an algorithm to iteratively update $(\bm{\mu},\bm{\beta})$ for the outer loop problem.
For notational simplicity, we define functions $\varphi_j(\beta_j)=\beta_j\Big(1+\exp\big(\hspace*{-0.5mm}-\hspace*{-0.5mm}a_j(P_{\mathrm{ER}_j}\hspace*{-0.5mm}-\hspace*{-0.5mm}b_j)\big)\Big)-M_j$
and $\varphi_{J+i}(\mu_i)=\mu_i\Big(1+\exp\big(\hspace*{-0.5mm}-\hspace*{-0.5mm}a_i(P_{\mathrm{ER}_i}\hspace*{-0.5mm}-\hspace*{-0.5mm}b_i)\big)\Big)-1$, $i\in\{1,\ldots,J\}$. It is shown in \cite{JR:sum_of_ratios,JR:sum_of_ratios1} that the unique optimal solution  $(\bm{\mu}^*,\bm{\beta}^*)$ is obtained if and only if  $\bm\varphi(\bm \mu,  \bm\beta)=[\varphi_1,\varphi_2,\ldots,\varphi_{2J}]=\zero$. Thus, the well-known damped Newton method can be employed to update $(\bm{\mu},\bm{\beta})$ iteratively. In particular, in the $n$-th iteration, ${\bm \mu}^{n+1}$ and ${\bm \beta}^{n+1}$ can be updated as, respectively,
\begin{eqnarray}\label{eqn:update_beta}
\hspace*{-2.5mm}{\bm \mu}^{n+1}\hspace*{-1.5mm}&=&\hspace*{-1.5mm}{\bm \mu}^{n}+\zeta^n\mathbf{q}^n\,\,\mbox{and}\,\  {\bm \beta}^{n+1}={\bm \beta}^{n}+\zeta^n\mathbf{q}^n,\\
\hspace*{-2.5mm}\mbox{where }\,\,\quad\mathbf{q}^n\hspace*{-1.5mm}&=&\hspace*{-1.5mm}[\bm\varphi'(\bm{\mu},\bm{\beta})]^{-1}\bm\varphi(\bm{\mu},\bm{\beta})
\end{eqnarray}
 and $\bm\varphi'(\bm{\mu},\bm{\beta})$ is the Jacobian matrix of $\bm\varphi(\bm{\mu},\bm{\beta})$. $\zeta^n$ is the largest  $\varepsilon^l$ satisfying
\begin{eqnarray}\label{eqn:convergence_condition}
\norm{\bm\varphi\big({\bm \mu}^{n}+\varepsilon^l\mathbf{q}^n,{\bm \beta}^{n}+\varepsilon^l\mathbf{q}^n\big)}\leq (1-\delta\varepsilon^l)\norm{\bm\varphi(\bm{\mu},\bm{\beta})},
\end{eqnarray}
where $l\in\{1,2,\ldots\}$, $\varepsilon^l\in(0,1)$, and $\delta\in(0,1)$. The damped Newton method converges to the unique solution $(\bm{\mu}^*,\bm{\beta}^*)$  satisfying the system of equations \eqref{eqn:conditions1} and \eqref{eqn:conditions2}, cf. \cite{JR:sum_of_ratios,JR:sum_of_ratios1}.
\begin{Remark}
{ We note that when there is one ER in the system, the traditional linear and the proposed non-linear EH model will lead to the same optimal resource allocation policy.}
\end{Remark}

\begin{Remark}
 { The signal model adopted in this paper  can be extended to include dedicated energy beams for the ERs by following a similar approach as in \cite{JR:Kwan_secure_imperfect}.}
\end{Remark}

\section{Results}
In this section, we present simulation results to demonstrate the system performance of the proposed resource allocation algorithm design.  We assume a carrier center frequency of $915$ MHz and a signal bandwidth of $200$ kHz.  There are $K=2$ IRs and $J$ ERs located $50$ meters and $10$ meters from the transmitter, respectively. Each ER is equipped with $N_{\mathrm{R}}=2$ receive antennas. {  Since the ERs are close to the transmitter, line-of-sight communication channels are expected. Hence, the multipath fading between the transmitter
and the ERs is modeled as Rician fading with a Rician factor of $3$
dB. In contrast, the IRs are located $50$ meters from the transmitter, thus, a line-of-sight may not be available and the multipath fading between the transmitter and the IRs is
modeled as Rayleigh fading.} All transmit antenna gains are $10$ dBi.  The thermal noise power  is $\sigma^2_{\mathrm{s}}=-95$ dBm. For the non-linear EH circuits, we set $M_j=20$ mW which corresponds to the maximum harvested power per ER. Besides, we adopt $a_j=6400$ and $b_j=0.003$ which were obtained by curve fitting for measurement data from  \cite{JR:EH_measurement_1}. The average system performance is obtained by averaging over different channel realizations.

\begin{figure}[t]
        \centering\vspace*{-5mm}
       \includegraphics[width=3.0 in]{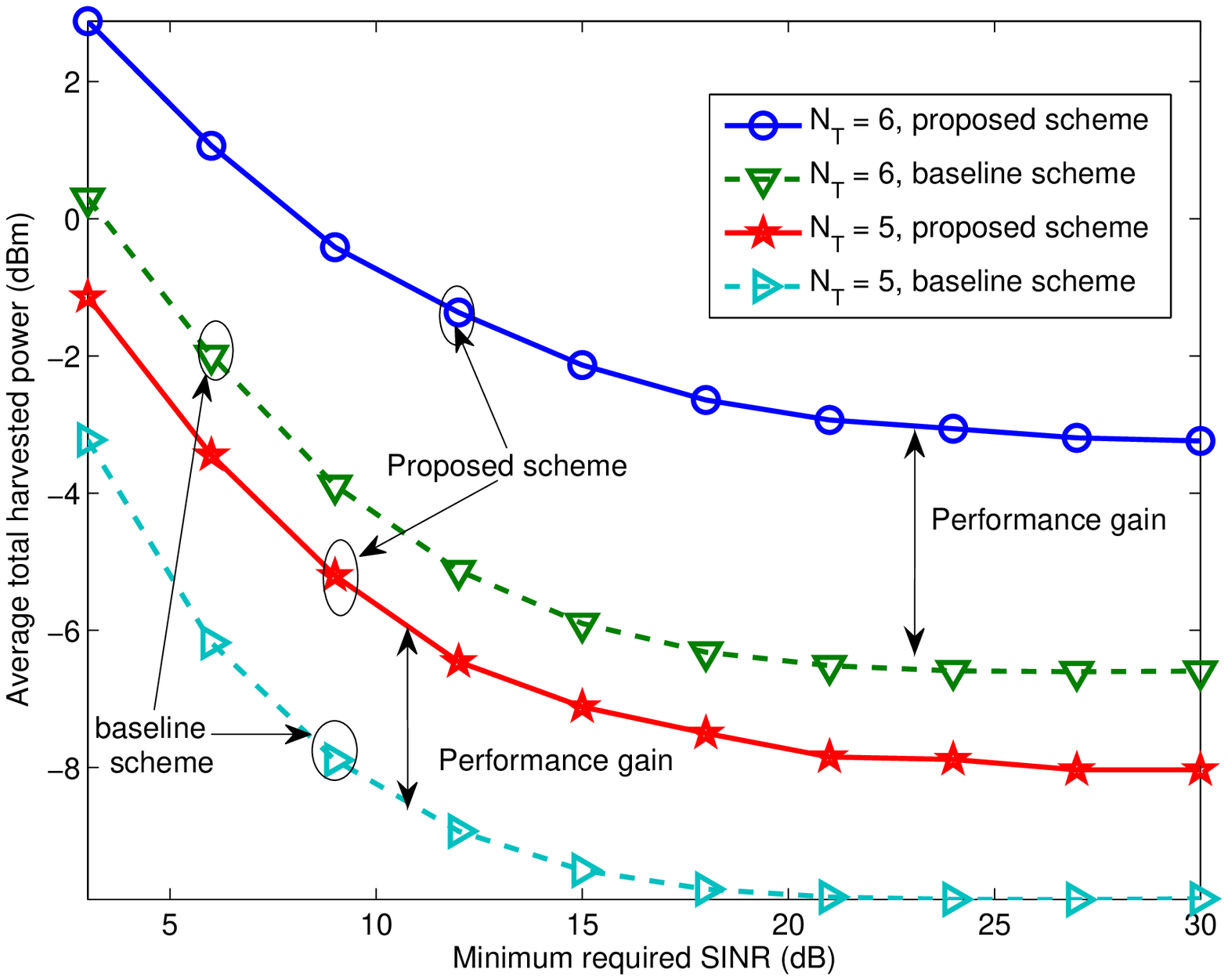}\vspace*{-3mm}
        \caption{Average total harvested power (dBm) versus the minimum required SINR (dB).}
        \label{fig:HP_SINR}
        \includegraphics[width=3.0 in]{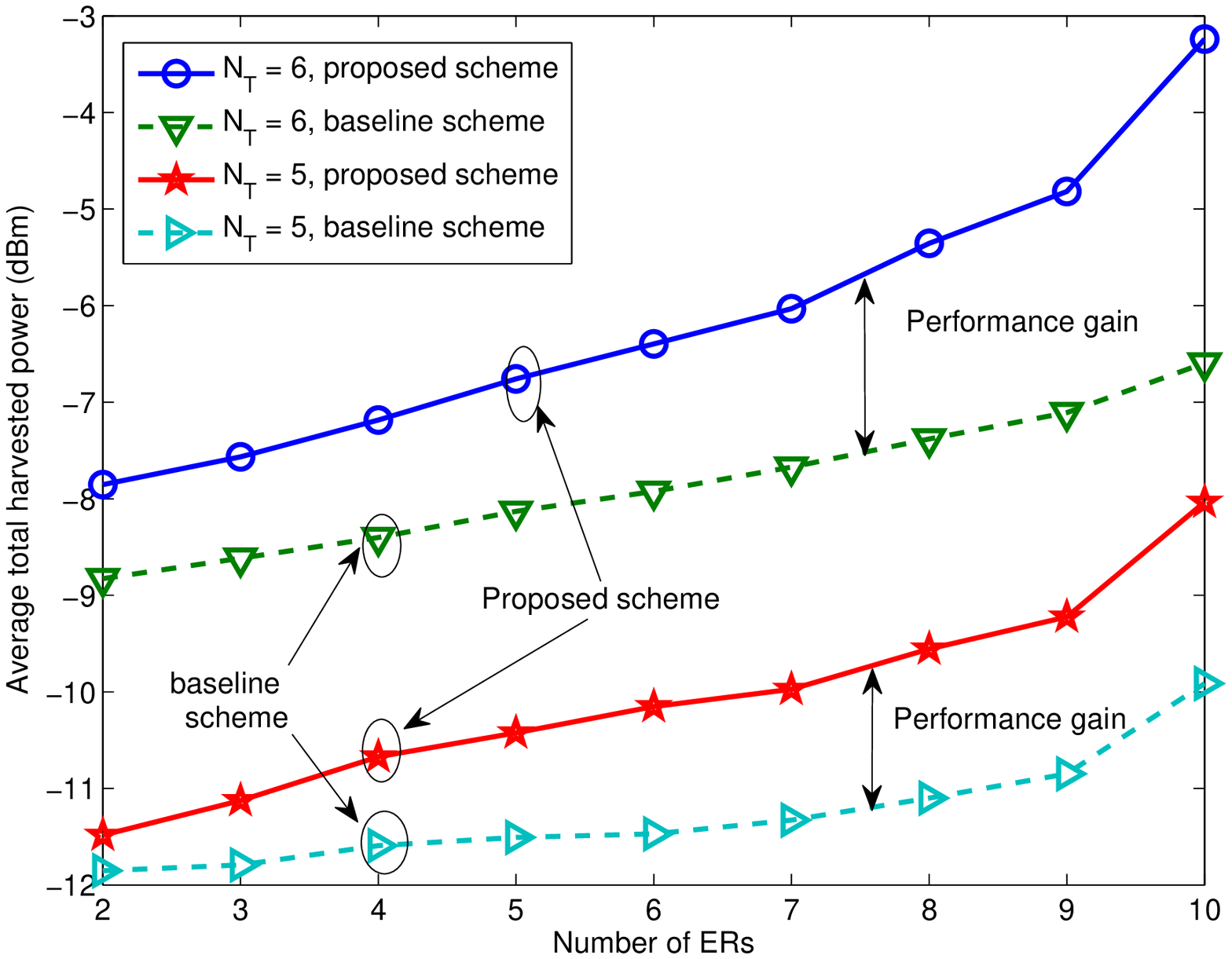}\vspace*{-3mm}
        \caption{Average total harvested power (dBm) versus the number of ERs.}
        \label{fig:HP_Pmax}\vspace*{-4mm}
\end{figure}
Figure \ref{fig:HP_SINR} depicts the average total harvested power versus the minimum received SINR at the IRs for $J=10$ ERs and different numbers of transmit antennas. We assume that all IRs require the same minimum receive SINR, i.e., $\Gamma^{\mathrm{req}}_k=\Gamma^{\mathrm{req}}$ and $P_{\max}=30$ dBm. { Extensive simulations (not shown here) have revealed that, in general, the proposed iterative algorithm converges to the globally optimal solution after less than $10$ iterations. } It can be observed from Figure \ref{fig:HP_SINR} that the average total harvested power is a monotonically decreasing function with respect to $\Gamma^{\mathrm{req}}$. Indeed, to satisfy a more stringent minimum SINR requirement, the transmitter is forced  to steer the direction of
transmission towards the IRs leading to a smaller amount of RF energy for
EH. On the other hand, the total harvested energy increases for an increasing number of transmit antennas $N_{\mathrm{T}}$, since the extra degrees of freedom offered by the increased number of transmit antennas facilitates a more power efficient resource allocation. For comparison, we also show the performance of a baseline
 scheme  in Figure
\ref{fig:HP_SINR}. For the baseline scheme, the resource allocation algorithm is optimized for maximization of the total system harvested power according to the conventional linear EH model in \eqref{eqn:linear_model} subject to constraints C1 and C2.
 As can be observed, the baseline scheme can only achieve a strictly smaller amount of total harvested power due to the resource allocation mismatch. In particular, the baseline scheme may cause saturation in EH in some ERs and underutilization of other ERs  because it does not account for the non-linear nature of the EH circuits.

Figure  \ref{fig:HP_Pmax} shows the average total harvested power versus the number of ERs $J$ for $P_{\max}=30$ dBm, a minimum required SINR of $30$ dB, and different numbers of transmit antennas $N_{\mathrm{T}}$. It can be observed that the average total harvested power increases with the number of ERs and the number of transmit antennas. In fact,   a larger portion of the radiated power can be
harvested when there are more ERs in the system since more
receivers participate in the EH process.
Besides, the performance gain of the proposed scheme compared to the baseline scheme increases with increasing number of ERs. This is because the resource allocation mismatch for the baseline scheme becomes more pronounced for a larger number of ERs  leading to unsatisfactory performance.
\section{Conclusions}\label{sect:conclusion}
In this letter, we proposed a practical EH model to capture the non-linear characteristics of EH circuits in SWIPT systems.  Furthermore, the resource allocation algorithm design for the proposed model was formulated as a
non-convex optimization problem with a sum-of-ratios objective function and was solved optimally by the proposed iterative algorithm. Our simulation results unveiled that resource allocation algorithms designed for the conventional linear EH model, which is widely used in the literature, may lead to resource allocation mismatches for practical non-linear EH circuits.

\section*{Appendix-Proof of Theorem \ref{thm:rankone}}\label{app:rankone}
It can be verified that  strong duality holds for the SDP relaxed version of \eqref{eqn:rank_constrained}. Thus,  solving the dual problem  is equivalent to solving the primal problem \cite{book:convex}. Now, we prove
Theorem \ref{thm:rankone} by first defining the Lagrangian function:
\begin{eqnarray} \label{eqn:appB7}
\hspace*{-3.5mm}{\cal L}\hspace*{-2.5mm}&=&\hspace*{-3.5mm}\sum_{k=1}^K\hspace*{-0.5mm}\Tr(\mathbf{B}_k\mathbf{W}_k)\hspace*{-0.5mm}+\hspace*{-0.5mm}
\sum_{k=1}^K\Tr\big((\mathbf{Y}_k\hspace*{-0.5mm}+\hspace*{-0.5mm}
\frac{\gamma_k\mathbf{H}_k}{\Gamma_{\mathrm{req}_k}})\mathbf{W}_k\big)\hspace*{-0.5mm}+\hspace*{-0.5mm}\Delta,\\
\hspace*{-3.5mm}\mathbf{B}_{k}\hspace*{-2.5mm}&=&\hspace*{-3.5mm}-\lambda\mathbf{I}_{N_{\mathrm{T}}}-\sum_{j \ne k}\gamma_j\mathbf{H}_j+\sum_{j=1}^J\rho_j\mathbf{G}_j\mathbf{G}_j^H,\label{eqn:A_k}
\end{eqnarray}
where $\Delta$ is the collection of all constants and variables that are independent of $\mathbf{W}_k$ and are thus not relevant in the proof. $\lambda,\gamma_k,\rho_j,$ and $\mathbf{Y}_k$ are the dual variables associated with constraints C1, C2, C4, and C5, respectively. Then, the dual problem of the SDP relaxed problem of \eqref{eqn:rank_constrained} is given by
\begin{eqnarray}\label{eqn:dual}
\hspace*{-1cm}\underset{ {\lambda,\gamma_k,\rho_j\ge0,\mathbf{Y}_k\succeq \mathbf{0}}}{\mino} \quad\underset{\mathbf{W}\in\mathbb{H}^{N_{\mathrm{T}}},\tau_j}{\maxo}\,\,\,{\cal
L}.\label{eqn:master_problem}
\end{eqnarray}
We define $\{\tau^*_j,\mathbf{W}_k^*\}$   and $\mathbf{\Xi}^*\triangleq\{\lambda^*,\gamma_k^*,\rho_j^*,\mathbf{Y}_k^*\}$ as the set of optimal primal and dual variables of the SDP relaxed version of \eqref{eqn:rank_constrained}, respectively. Now, we consider the following Karush-Kuhn-Tucker (KKT) conditions:
\begin{eqnarray}
\hspace*{-3mm}\mathbf{Y}_k^*\hspace*{-3mm}&\succeq&\hspace*{-3mm}\mathbf{0},\,\,\lambda^*,\gamma_k^*,\rho_j^*\ge 0,\,\forall k,\,\forall j, \label{eqn:dual_variables}\\
\hspace*{-3mm}\mathbf{Y}_k^*\mathbf{W}_k^*\hspace*{-3mm}&=&\hspace*{-3mm}\mathbf{0},\label{eqn:KKT-complementarity}\\
\hspace*{-3mm}\mathbf{Y}_k^*\hspace*{-3mm}&=&\hspace*{-3mm}-\mathbf{B}_{k}^*-\frac{\gamma_k^*\mathbf{H}_k}{\Gamma^{\mathrm{req}}_k}, \label{eqn:lagrangian_gradient}
\end{eqnarray}
where  $\mathbf{B}_{k}^*$ is obtained by substituting the optimal dual variables $\mathbf{\Xi}^*$ into (\ref{eqn:A_k}). {
Equation (\ref{eqn:KKT-complementarity})
 is the complementary slackness condition which is obtained by taking the derivative of the Lagrangian function with respect to $\mathbf{Y}_k^*$. Besides, equation (\ref{eqn:KKT-complementarity}) indicates that  the columns of $\mathbf{W}^*_k$ lie in the null space of $\mathbf{Y}^*_k$ for $\mathbf{W}^*_k\ne\mathbf{0}$.}
 Therefore, if  $\Rank(\mathbf{Y}^*_k)=N_{\mathrm{T}}-1$, then the optimal beamforming matrix is a rank-one matrix. To reveal the structure of $\mathbf{Y}^*_k$, we show by contradiction that $\mathbf{B}_k^*$ is a negative definite matrix with probability one. For a given set of optimal dual variables, $\mathbf{\Xi}^*$, and $\tau_j^*$,    (\ref{eqn:dual}) can be written as
\begin{eqnarray}\hspace*{-2mm}\label{eqn:dual2}
\,\,\underset{\mathbf{W}_k\in\mathbb{H}^{N_{\mathrm{T}}}}{\maxo} \,\, {\cal L}.
\end{eqnarray}
Suppose $\mathbf{B}_k^*$ is not negative definite, then we can construct $\mathbf{W}_k=r\mathbf{v}_k\mathbf{v}_k^H$ as one of the optimal solutions of (\ref{eqn:dual2}), where $r>0$ is a scaling parameter and $\mathbf{v}_k$ is the eigenvector corresponding to one of the non-negative eigenvalues of $\mathbf{B}_k^*$.  We substitute $\mathbf{W}_k=r\mathbf{v}_k\mathbf{v}_k^H$ into (\ref{eqn:dual2}) which leads to
${\cal L}=\sum_{k=1}^K\Tr(r\mathbf{B}_k^*\mathbf{v}_k\mathbf{v}_k^H)+r\sum_{k=1}^K\Tr\Big(\mathbf{v}_k\mathbf{v}_k^H
\big(\mathbf{Y}_k^*+\frac{\gamma_k^*\mathbf{H}_k}{\Gamma^{\mathrm{req}}_k}\big)\Big)+\Delta$. Since the channels of $\mathbf{G}_j$ and $\mathbf{h}_k$ are assumed to be statistically independent, it can be shown that $\gamma_k^*>0$ for the optimal solution. Also, it follows that by setting $r\rightarrow \infty$, the dual optimal value  becomes unbounded from above. However, the optimal value of the primal problem is finite for a finite $P_{\max}$. Thus,  strong duality does not hold which leads to a contradiction. Therefore, $\mathbf{B}_k^*$ is a negative definite matrix with probability one, i.e., $\Rank(\mathbf{B}_k^*)=N_{\mathrm{T}}$. Then, by exploiting (\ref{eqn:lagrangian_gradient}) and a basic  inequality for the rank of matrices, we have
\begin{eqnarray}\notag
\hspace*{-3mm}&&\hspace*{-2mm}\Rank(-\mathbf{Y}^*_k)=\Rank(\mathbf{Y}^*_k)= \Rank\Big(\mathbf{B}_k^*+\frac{\gamma_k^*\mathbf{H}_k}{\Gamma^{\mathrm{req}}_k}\Big)\\
\hspace*{-3mm}&\ge&\hspace*{-2mm}
 \Rank(\mathbf{B}_k^*)-\Rank\Big(\frac{\gamma_k^*\mathbf{H}_k}{\Gamma^{\mathrm{req}}_k}\Big) = N_{\mathrm{T}}-1.
\end{eqnarray}
Furthermore, $\mathbf{W}_k^*\ne\mathbf{0}$ is required to satisfy the minimum SINR requirement of  IR $k$ in C2 for $\Gamma^{\mathrm{req}}_k>0$. Hence, $\Rank(\mathbf{Y}^*_k)=N_{\mathrm{T}}-1$ and $\Rank(\mathbf{W}^*_k)=1$. \qed

\bibliographystyle{IEEEtran}
\bibliography{vicky_leng}

\end{document}